# Structural modifications of GeO$_2$ glass under high pressure and high temperature

Antoine Cornet,[1,a)] Rémy Molherac,[1] Bernard Champagnon,[1] and Christine Martinet[1]

[1]*Institut Lumière Matière, UMR5306 Université Lyon1-CNRS, Université de Lyon 69622 Villeurbanne cedex, France*

Vitreous GeO$_2$ has been compressed at high temperature, to investigate the effect of thermal activation on the structural reorganization during compression. The measurements were performed in-situ using micro Raman spectroscopy under pressure up to 6 GPa and temperature up to 400°C. The evolution of the Raman shift of the main band (400-500 cm$^{-1}$) with temperature during compression evidences a pressure window around 3 GPa within which temperature has a remarkable influence on the structure, in particular the intermediate range order. We find that this change is well correlated with previous ex-situ density measurements from high pressure-high temperature densifications. Moreover, coordination changes from tetrahedrally (GeO$_4$) to octahedrally (GeO$_6$) coordinated GeO$_2$ are accelerated with the heating during the compression.

[a)] Author to whom correspondence should be addressed.  Electronic mail:  antoine.cornet@univ-lyon1.fr

## I. Introduction

Because they are considered as archetypal glasses, vitreous silica and germania remain a preferential test material for fundamental studies on the glassy state. In earlier studies,[1,2] it has been demonstrated that when compressing these glasses at high temperature, it significantly decreases the elastic limit, i.e. the pressure threshold for permanent densification. Hence, there has been an increasing interest in understanding how a combined high-temperature (HT) and high-pressure (HP) of silica or germania affects the structural evolution of these glasses during compression. A pressure-temperature map of silica has been established by Inamura based on x-ray diffraction measurements,[3] in which a region of temperature-pressure coupling has been identified. Thus, a new theoretical model was needed to explain this kind of densification. In this regard, the rigidity percolation model developed at the same time by Trachenko et al.[4,5] provides predictions that are nicely consistent with experimental data.

In the case of glassy germania, Shen et al.[6] studied the plastic domain, where the transition from tetrahedral $GeO_4$ units to octahedral $GeO_6$ units has already been initiated as a result of the applied pressure. Using x-ray diffraction, they evidenced that heating enhances the coordination change, inducing a larger densification ratio compared to cold compression for the same maximum pressure. However, subjecting the glass to high temperature cannot initiate the $GeO_4$ to $GeO_6$ transformation.[6] If pressure is not sufficient to provoke this change, temperature affects the medium range order only. In this study, we specifically investigate the impact of high temperature on a compressed $GeO_2$ glass. We focus on the intermediate range order structure revealed by in-situ Raman spectroscopy experiments and we discuss about the coordination change.

## II. Experimental

Pieces of $GeO_2$ glass of about 50*50*30 $\mu m^3$ were introduced in a membrane-driven Diamond Anvil Cell (DAC) equipped with an external resistive heating. The diamond culet size is 400 μm. The gasket consisted of 200 μm thick disks of inox indented and drilled to form a sample chamber of 40 μm in thickness and 150 μm in diameter. Ruby and YAG:$Sm^{3+}$ chips were introduced in the sample chamber in order to measure both *in-situ* pressure and temperature. The transmitting medium was Argon introduced using a cryogenic method, a fully hydrostatic medium in the pressure range covered in our experiments.[7] The emission of $^2E$-$^4A_2$ (R1 line) $Cr^{3+}$ transition in the ruby has been well established as a function of pressure.[8] YAG:$Sm^{3+}$ has been proposed as a pressure sensor for high temperature experiments because it exhibits an intense luminescence doublet Y1 at 16185 $cm^{-1}$ and Y2 at 16231 $cm^{-1}$ almost insensitive to temperature.[9] The Y1 line in the emission spectrum of YAG:$Sm^{3+}$ has been calibrated in house, as values found in the literature vary.[9-11] For more precisions, we then calibrated the evolution of the Y1 and Y2 peaks position under pressure and temperature as described elsewhere.[12] We also calibrated the temperature dependence of the luminescence spectra of $Cr^{3+}$ in the ruby. These calibrations can be described by the following equations.

$$P_{YAG}\ (GPa) = -2.10^{-4} \times (\varpi - \varpi_0)^2 + 0{,}142 \times (\varpi - \varpi_0) \quad (1)$$

$$\left(\frac{\partial T}{\partial \varpi}\right)_{Ruby} = 6{,}434\ K.cm^{-1}, \quad (2)$$

where $P_{YAG}$ is the pressure in GPa obtained from the YAG:$Sm^{3+}$ emission spectrum, $\varpi$ and $\varpi_0$ are the frequencies in $cm^{-1}$ of the Y1 emission of $Sm^{3+}$ in YAG:$Sm^{3+}$ at that pressure and at ambient pressure, respectively. Equation 2 describes the R1 line emission of $Cr^{3+}$ in ruby as a function of temperature and this relation is independent of the applied pressure, while the frequency of the Y1 line of YAG:$Sm^{3+}$ does not change with temperature. Thus, YAG:$Sm^{3+}$ is perfectly suitable as a pressure sensor for HP-HT experiments. Similarly, the shift of the emission of $^2E$-$^4A_2$ ($R_1$ line) $Cr^{3+}$ transition in the ruby is linear as a function of temperature at all pressures, and can therefore be reliably used to determine the temperature during the (P,T) experiments.

The first HP-HT runs were realized with an additional thermocouple located near the diamonds. The difference between the temperatures measured near the diamonds and deduced from $Cr^{3+}$ ruby emission showed differences smaller than 15°C for the full 25°-450°C temperature range. Thus, the fluorescence technique is the best way to get both pressure and temperature values since they allow measurements inside the sample chamber, i.e. close to $GeO_2$ glass. The Raman spectra were recorded using a Horiba Jobin Yvon Aramis spectrometer configured with a 1800 lines per mm grating. The DAC was set up in a back scattering configuration under a UHWD x20 Olympus objective. The excitation occurs at the 473 nm line of a diode pump laser, with an incident power of 15 mW. The typical acquisition times are about ten minutes. From each spectrum we subtract a baseline that consists of the Raman spectrum collected of the pressure medium close to the sample in the DAC chamber and acquired in the same conditions.

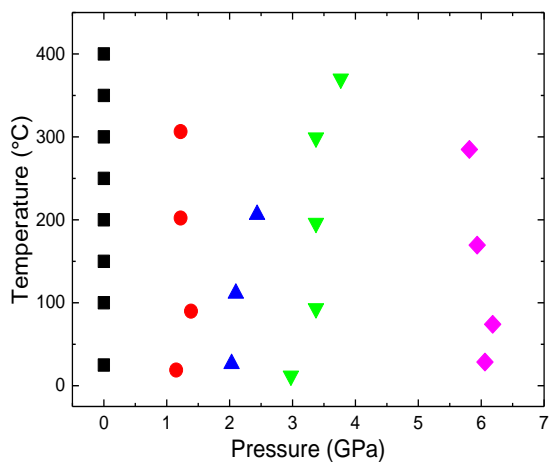

FIG. 1. Pressure temperature map of our experiments. The pressure and temperature errors on each value are about 0.3 GPa and 15°C.

To investigate the influence of temperature on the structure, pressure has to be kept constant during the measurement. We first establish the desired pressure, and then increase the temperature in steps of 50°C or 100°C. During heating, the pressure is adjusted continuously so as to maintain the desired value. Because this was not strictly the case, we report mean values of the pressure and associated uncertainties for each run (maximally 0.5 GPa). Once the target temperature value is achieved, we allow one hour for temperature and pressure to stabilize. As illustrated in Fig. 1, we performed four HP-HT runs at 1.4, 2.2, 3.4, 6.0 GPa and one run at the atmospheric pressure.

## III. Results

All the experimental (P,T) conditions, for which Raman measurements have been performed, are summarized in Fig. 1. The in-situ Raman spectra obtained at 3.4 and 6.0 GPa and at different temperatures are shown in Fig. 2.

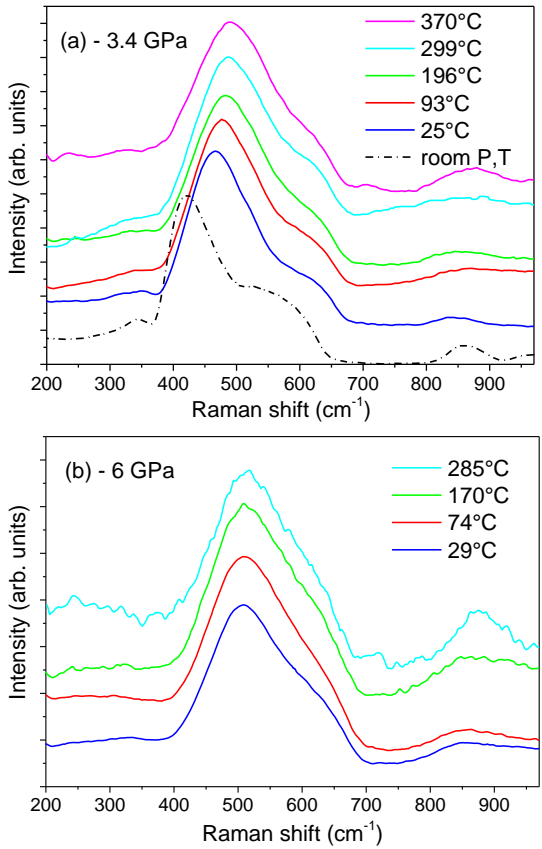

FIG. 2. Raman spectra of $GeO_2$ glass compressed at (a) 3.4 GPa and (b) 6.0 GPa at various temperatures. The dashed spectrum on the Fig. 2 (a) corresponds to non-densified $GeO_2$ glass at ambient conditions of pressure and temperature.

The dashed line in Fig. 2(a) corresponds to the spectrum of non-densified amorphous $GeO_2$ under atmospheric conditions. In this spectrum, the main band at 420 cm$^{-1}$ and the band centred at 850 cm$^{-1}$ are related to the bending and asymmetric stretching modes of the Ge-O-Ge bond, respectively.[13] Thus, the intensity of these bands are related to the degree of connectivity between adjacent tetrahedron, and thus provide information about the intermediate range order. The shoulder in the high frequency part of the main band (MB) at 420 cm$^{-1}$ can be deconvoluted into three components.[14] In the following, we focus on the behavior of this main band.

The positional shift of the intensity maximum of the main band with temperature, as can be seen in figures 2a and 2b, between ambient and nearly 400°C, and for either pressure, is roughly one order of magnitude weaker than that caused by varying pressure, e.g., comparing the spectra at ambient conditions and 6 GPa. The main band shifts slightly toward high frequencies at 3.4 GPa with increasing temperature. At a fixed pressure, both at 3.4 GPa and 6.0 GPa, the width of the main band and its shoulder tend to increase with temperature. At the highest temperature, we also observe the appearance of a band at 700 cm$^{-1}$. In the spectra for samples heated while subject to 6.0 GPa of applied pressure, we notice the appearance of broad-ranged spectral features around 280 cm$^{-1}$.

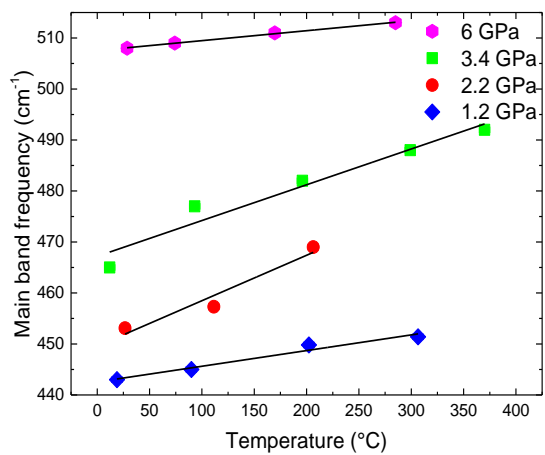

FIG. 3. In situ main band frequency against temperature at various pressures. The slopes give the quantity $dv_{MB}/dT$.

The positions of the main band (MB) for each temperature at different pressures are plotted in Fig. 3. At any given pressure, the MB position as a function of temperature exhibits a linear behavior with a positive slope, $dv_{MB}/dT$, indicating a reduction of the average Ge-O-Ge angle associated with the network compaction.[15] Given that the slopes observed at 2.2 and 3.4 GPa are larger than those at 1.2 and

6.0 GPa, the trend in the temperature dependence of the MB shift as a function of the applied pressure, i.e., $d\nu_{MB}/dT$ vs. P, is not monotonous.

**IV. Discussion**

It is known from the very first studies on $GeO_2$[2] and $SiO_2$[1, 2, 16] glasses that subjecting them to high temperatures during compression leads to permanent densification at pressures lower than would be required at room temperature to achieve the same result. The commonly accepted scheme for cold compression of $GeO_2$ and $SiO_2$ glasses consist of two or three steps[17]. First, a homogenization of the network for silica, apparent in the reduction of the boson peak[18, 19, 20] and the decrease in the intertetrahedral angle distribution, these processes occur within the regime of elastic deformations. The compaction of the network progresses until the cation-cation repulsion becomes too strong, which constitutes the elastic limit. Above this limit, one observes a progressive increase of Ge or Si coordination number,[21] e.g., from 4 to 6 during compression. The breaking and the redistribution of the network bonds allows for the structure to relax towards a more densely packed configuration. Recent and carefully conducted x-ray and neutron diffraction in-situ experiments on $SiO_2$ and isotopically enriched $GeO_2$ glass gives clear quantitative evidence for this structural evolution process, revealing the reduction of the Ge-O-Ge bond angle and the increasing cation coordination across consecutive pressure ranges.[21] The Ge or Si coordination returns to 4 at room pressure for silica and germania.

The fact that heating during compression reduces the elastic limit can be interpreted as either due to enhanced thermal activation of the structural evolution mechanisms described above, or as an indication for an altogether different structural modification route. Several molecular dynamics studies examining the mechanisms of pressure-induced densification conclude that the redistribution of intertetrahedral bonds within the structure is a thermally activated process, while varying pressure thresholds are reported concerning the apparition of higher coordinated Si or Ge atoms.[5, 22] Moreover, the bond redistribution seems to be an essential aspect of the relaxation at the intermediate range order.

To further develop our understanding of the thermally assisted densification of glasses, it is beneficial to discuss our results in the context of previous results on silica and make an analogy between silica and germania. Concerning silica, the existence of a pressure window for high-temperature induced fast densification has been identified using in situ x-rays diffraction experiments and ex-situ density measurements. On a pressure-temperature map up to 1000°C and 20 GPa, Inamura et al.[3] delineated the area where the shape and position of the first sharp diffraction peak (FSDP) undergoes a significant change. They found that the maximum effect on the intermediate range order is obtained at about 5 GPa. This pressure is consistent with the ex-situ density measurements performed previously,[23,24] and with the molecular dynamic simulations by Trachenko.[5] All these studies show that the effect of temperature on the density is maximal around 5 GPa, which corresponds approximately to half the pressure of the room temperature elastic limit, i.e. 9 GPa.

Shen et al.[6] performed a similar diffraction study on $GeO_2$ glass at different temperatures and showed unambiguously that over-coordinated Ge atoms are absent below the elastic limit at room temperature, i.e. 4 GPa. Moreover, as shown in Fig. 2(a), the different Raman spectra recorded below 4 GPa do not exhibit any growth of a band located at 280 cm$^{-1}$, which Durben et al. have linked to the Ge coordination increase.[17] Thus, temperature induced densification in this pressure range is associated with relaxations in the intermediate range structure, which possibly affects the boson peak, the ring statistics, and the Ge-O-Ge inter-tetrahedral angle distribution. Here, we focus on Ge-O-Ge angles. The rates of MB frequency change with temperature, i.e., $dv_{MB}/dT$ (Fig. 3), are plotted as a function of pressure in Fig. 4. In this figure, we also included the ratio of recovered densities from hot and cold compressions. For that, density data from Stone et al. of non-densified $GeO_2$ glass and of $GeO_2$ glass that was compressed at 400°C to 2.0 GPa, 3.0 GPa, 6.0 GPa [25], and density data from cold compression from Richet et al.[26] where used.

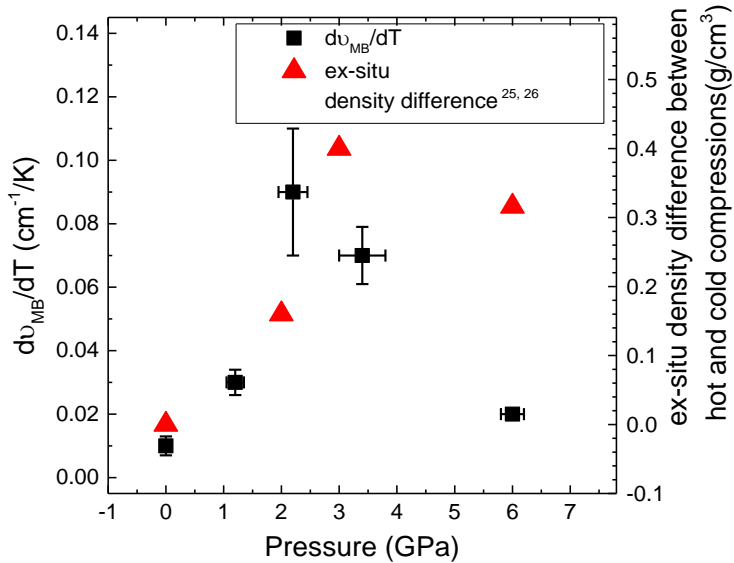

FIG. 4. Rates of main band frequency change with temperature ($d\nu_{MB}/dT$) as a function of the applied pressure measured in this study (squares ■); Difference between recovered $GeO_2$ glass densities compressed at room temperature and 400°C to 2.0, 3.0 and 6.0 GPa, (triangles).[25,26].

The $d\nu_{MB}/dT$ slopes display a maximum at around 3 GPa, which demonstrates that the impact of temperature on the structure during compression is only significant within a limited pressure window. Moreover, the maximum impact due to temperature is reached for pressures between 2-4 GPa. This pressure range corresponds to an elastic domain for cold compression i.e. at room temperature, where no irreversible bond breaking appears. There is a direct link between the structural process responsible for the main band shift and the overall density. The local environment after the bond redistribution that permits denser atomic packing is characterized by a reduction in the Ge-O-Ge angle that causes a shift of the main band position to high frequencies, which in turn is an accurate marker for this process. Trachenko et al. have shown in silica glass that the in situ volume ratio between hot and cold compression follows the same behaviour as the $d\nu_{MB}/dT$ quantities presented in Fig. 4, i.e., a bell-shaped curve centred approximately at a pressure corresponding to half of the elastic limit value at room temperature, i.e. 5 GPa [5].

In Fig. 4 we compare the rates of MB frequency changes with temperature to the difference between recovered GeO$_2$ glass densities compressed at room temperature and 400°C to 2.0, 3.0 and 6.0 GPa.[25,26] For as long as dv$_{MB}$/dT is independent of temperature, this differential can be directly mapped onto an MB frequency difference over an arbitrary temperature range, and the comparison in Fig.4 permits us to analyse the relationship between the MB shift, or the Ge-O-Ge bond angle change, and the degree of permanent densification. Below 4 GPa, in the domain of pressure where no permanent densification is possible at room temperature, we find a good correlation between our present values of dv$_{MB}$/dT and the difference between recovered densities upon cold or hot compression. Accordingly, in this regime, densification that can be attributed to Ge-O-Ge bond angle changes require a heat treatment in order to stabilize. Based on the density data by Stone for compression at 400°C (shown in Fig. 4),[25] bond angle stabilization can lead up to 70% irreversible densification in GeO$_2$ for pressures as low as 3-4 GPa.

In Fig. 4, we see that the value of dv$_{MB}$/dT is much reduced at 6.0 GPa, but the difference in recovered densities between hot and cold compression is still significant. It demonstrates that once the room temperature elastic limit is exceeded, i.e. above 4 GPa, the coordination change is initiated and the change in the inter-tetrahedral angle plays a comparably insignificant role in the permanent densification. Moreover, the transformation from a tetrahedral to octahedral network one is confirmed by the growth of the band located around 280 cm$^{-1}$.[17] This is consistent with the study of Shen[6] who demonstrated that once the coordination change from tetrahedron to octahedron is initiated via applied pressures exceeding 4 GPa, it becomes the dominant densification mechanism induced by an increase of temperature.

The weak band appearing at 700 cm$^{-1}$ on the spectra recorded at the highest temperatures is attributed to the A$_{1g}$ mode in the crystalline rutile like structure.[14] This band is the strongest in the Raman spectrum of the crystalline rutile polymorph of GeO$_2$, thus we consider it to be a crystallization precursor. The other Raman active modes in the rutile like polymorph of GeO$_2$ are weaker and are at 173 and 873 cm$^{-1}$ (see ref. 14). So, in the region of interest, i.e. 300-700 cm$^{-1}$, the Raman spectrum arises from the glass only, and a beginning crystallization will not interfere in the results presented here.

**V. Conclusion**

We studied the evolution of glassy GeO$_2$ by exploring a HP-HT map between ambient and 6 GPa and 25-400°C through in-situ micro-Raman spectroscopy. We identified the existence of a pressure range where the temperature influence on the structure during compression is maximal. In this pressure range, located between 2 GPa and 3 GPa, heat is required activate structural relaxation that stabilizes the pressure-induced bond angle changes, and to cause permanent densification. Indeed, the structural changes observed by Raman spectroscopy are in good correlation with previous data on ex-situ density for GeO$_2$ recovered samples from HP-HT compression. At pressures exceeding 4 GPa, we found that heat boosts the network former coordination increase, bond angle adjustments play an insignificant role.

**Acknowledgments**

The authors are thankful to the vibrational spectroscopies Platform at University Lyon 1 France (CECOMO) facility for micro-Raman spectroscopy experiments. The authors would like to thank Prof. John Kieffer for helpful discussions.

[1] P.W. Bridgman and I. Simon, J. Appl. Phys. **24**, 405 (1953)

[2] J.D. Mackenzie, J. Am. Ceramic Society **10**, 461470 (1963)

[3] Y. Inamura, Y. Katayama, W. Utsumi and K.I. Funakoshi, Phys. Rev. Lett. **93**, 015501 (2004)

[4] K. Trachenko, M.T. Dove, Phys. Rev. B. **67**, 064107 (2003)

[5] K. Trachenko, M.T. Dove, V. Brazhkin and F.S. El'kin, Phys. Rev. Lett. **93**, 13502 (2004)

[6]G. Shen, H. Liermann, S. Sinogeikin, W. Yang, X. Hong, C. Yoo and H. Cynn, PNAS **104**, 14576 (2007)

[7]S. Klotz, J.-C. Chervin, P. Munsch and G. Le Marchand, J. of Phys. D : App. Phys. **42**, 075413 (2009)

[8]Chervin J C, Canny B and M. Mancinelli, High Pressure Res. **21,** 305-314 (2001)

[9]C. Sanchez-Valle, I. Daniel, B. Reynard, R. Abraham, C. Goutaudier, J.Appl. Phys., **92**, 4349 (2002)

[10]N. J. Hess, G. J. Hexharhos, High Press. Res., **2**, 57 (1989)

[11]N. J. Hess, D. Schiferl, J. Appl. Phys., **68**, 1953-1960 (1990)

[12]J. Lam, A. Chemin, C. Martinet, A. Cornet, K. Lebbou, C. Dujardin, G. Ledoux, F. Chaput, B. Gökce, S. Barcikowski and D. Amans (to be submitted)

[13]L. Giacomazzi, P. Umari and A. Pasquarello, Phys. Rev. Lett. **95**, 075505 (2005)

[14]M. Micoulaut, L. Cormier and G.S. Henderson, J. Phys.: Condens. Matter **18**, 753-784 (2006)

[15]P.N. Sen and M.F. Thorpe, Phys. Rev. B. **15**, 4030 (1977)

[16]P. McMillan, B. Piriou and R. Couty, J. CHem. Phys. **81**, 4234-4236 (1984)

[17]D.J. Durben, G.H. Wolf, Phys. Rev. B **43**, 2355-2362 (1991)

[18]T. Deschamps, Ph.D thesis, Université Claude Bernard Lyon 1, 2009

[19]J. Schroeder, W. Wu, J.L. Apkarian, M. Lee, L. Hwa and C.T. Moynihan, J. of Non-Cryst. Solids **349**, 88-97 (2004)

[20]R.J. Hemley, C. Meade, H. Ho-Kwang, Phys. Rev. Lett. **79**, 1420 (1997)

[21]P.S. Salmon and A. Zeidler, J. Phys.: Condens. Matter **27**, 133201 (2015)

[22]L. Huang and J. Kieffer, Phys. Rev. B **69**, 224204 (2004)

[23]C. Martinet, A. Kassir-Bodon, T. Deschamps, A. Cornet, S. Le Floch, V. Martinez, B. Champagnon, J. Phys. : Condens. Matter **27**, 325401 (2015)

[24]B.T. Poe, C. Romano, G. Henderson, J. Non-Cryst. Solids **341**, 162-169 (2004)

[25]C.E. Stone, A.C. Hanon, T. Ishihara, N. Kitamura, Y. Shirakawa, R.N. Sinclair, N. Umesaki and A.C. Wright, J. of Non-Cryst. Solids **293**, 769-775 (2001)

[26]P. Richet, G. Hovis and B. Poe, Chemical Geology **213**, 41-47 (2004)